\documentclass[3p,times,twocolumn]{elsarticle}

\journal{Special Issue on Data Distribution in Industrial and Pervasive Internet}

\def\BibTeX{{\rm B\kern-.05em{\sc i\kern-.025em b}\kern-.08em T\kern-.1667em\lower.7ex\hbox{E}\kern-.125emX}}

\usepackage{times}
\usepackage{gensymb}
\usepackage{tabularx}
\usepackage{lipsum}
\usepackage{array}
\usepackage{booktabs}
\usepackage{listings}
 
\usepackage{enumerate}
\usepackage{enumitem}

\usepackage{multirow}

\usepackage[caption]{subfig}

\lstdefinestyle{lststyle}{
 captionpos=b, 
 tabsize=2,
 basicstyle=\linespread{0.9}\footnotesize\ttfamily,
}
 
\begin{document}

\begin{frontmatter}

\title{A Scalable IoT-Fog Framework for Urban Sound Sensing}

\author[mymainaddress]{Marc Jayson Baucas}
\author[mymainaddress]{Petros Spachos\corref{mycorrespondingauthor}}
\cortext[mycorrespondingauthor]{Corresponding author}
\ead{petros@uoguelph.ca}

\address[mymainaddress]{School of Engineering, University of Guelph, Guelph, Ontario, Canada}

\begin{abstract}
 \noindent Internet of Things (IoT) is a system of interrelated devices that can be used to allow large-scale collection and analysis of data. However, as it grew,  IoT networks were not capable of managing the data from these services. As a result, cloud computing was introduced to address the need for datacentres for IoT networks. As the technology evolved, the demand for a proper means of supporting and managing crowdsensing and real-time data increased, and cloud servers could no longer keep up with the large volumes of incoming data. This demand brought rise to fog computing. It became an extension to the cloud and allowed resources to be allocated around the network effectively. Its integration to IoT reduced the strain towards the cloud servers. However, issues in high power consumption at the end device and data management constraints surfaced. This paper proposes two approaches to alleviate  these issues to keep fog computing remain as a reliable option for IoT-related applications. We created an IoT-based sensing framework that used an urban sound classification model. Through active low and high power states and resource reallocation, we created a network configuration. We tested this configuration against IoT frameworks that use the default fog and cloud setups. The results improved the framework's end device power consumption and server latency. Overall, with the proposed framework, fog computing was proven to be capable of supporting a  scalable IoT  framework for urban sound sensing.   
\end{abstract}

\end{frontmatter}

\section{Introduction}
Internet of Things (IoT) network is a platform that allows devices to communicate with one another via a wireless connection~\cite{iot-crit}. As a result, industries have developed many promising applications based on IoT~\cite{dxu}. Some examples of these applications are mobile asset tracking, micro-location, secure communication, and environment sensing~\cite{yaq, spachos, iot-applications}. With the emergence of IoT development, large volumes of data are being generated by IoT devices every day~\cite{iot-data}. This growth shows the need for dedicated storage and more processing capacity, which resulted in the introduction of cloud computing. It is a model that implements an on-demand provider of computing and online storage services~\cite{cloud-broker, iot-cloud, caval,botta}. 

With the use of cloud computing in an IoT network, devices can access software applications and infrastructure without the need of owning them~\cite{iot-cloud}. Industries used cloud computing to open IoT-related services and applications that have larger scopes. Some of these applications are in home automation, healthcare, and navigation~\cite{iot-cloud-services}.

However, cloud services are now running into the same issues that IoT networks had before~\cite{iot-fog-sensing, iot-fog-enabled}. As networks expanded, so did the demand for its services. Too many devices are attempting to connect and request data from the server at the same time. As a result, data travels in and out of a server at increased rates. Cloud-based IoT networks used for  centralized data management and storage~\cite{cloud-smart}. However, when it comes to scaling networks, one server cannot handle a network of end devices that are continuously growing~\cite{big-data-limits}. Applications that require user mobility, low latency, and location awareness becomes too hard for a centralized cloud-based IoT network. As a result, more complex real-time data processing and crowdsensing services become more of a pipedream with cloud-based IoT solutions~\cite{iot-fog-enabled}. 

As a means of addressing this issue, fog computing became an extension to cloud computing~\cite{peng, fog-priv}. Fog computing is a process allocation technique that offloads the different sections of the program from the server to the end devices. Offloading creates a better means of balancing the number of computations between the fog and the cloud~\cite{iot-fog-sensing}.

This paper presents two approaches to improve device energy efficiency and server load balancing in an IoT network. These approaches are active low and high power state, and process reallocation. Focusing on these two can help our proposed IoT framework become more scalable. We selected environmental sensing for our implementation. Environmental sensing is a general topic, hence, we focused on one of its subsets, which is urban sound classification~\cite{Urbansound}. 

This paper built on our previous work in~\cite{edgeiot}, which focuses more on the comparison between cloud-based and fog-based IoT networks. We point out the importance of process optimization towards the realization of a configuration that can create an efficient framework. On the other hand, this paper focuses more on the scalability of the framework. We add more tests in an attempt to isolate other key factors that might contribute to improving the framework's scalability.

The rest of this paper is as follows: Section~\ref{bground} is an overview of fog computing and an introduction to our focused issues and proposed solutions. Section~\ref{proposal} discusses the design specifications of our framework. Section~\ref{exp} introduces the experimental setup leading to the conducted tests and results. Finally, Section~\ref{conc} are the conclusions to the study. 

\section{Fog Computing Background} \label{bground}
This section provides an overview of fog computing as an extension of cloud computing in IoT networks. Also, it discusses the different fog computing categories and which one are we using in our framework.

\subsection{Overview of Fog Computing with IoT}
IoT networks lacked scalability in a growing network of devices~\cite{fog-app}. Cloud computing enabled IoT to handle the incoming wave of large-scale applications ~\cite{iot-cloud}. However, as the market for IoT-cloud services evolved, the demand for new and innovative applications and services arose. Examples of such as smart grids and smart homes pushed IoT and the cloud to its limits~\cite{cloud-smart}. Most services now demand applications with features such as real-time sensing, larger capacities, which cloud servers are no longer able to sustain~\cite{fog-priv}. As a result, fog computing became an extension of cloud computing. Also, known as clouds at the edge, it uses nodes within the boundaries of the network to carry out computations~\cite{fog-alloc}. Fog computing is a computing paradigm that integrates with the cloud infrastructure. It creates a computing facility for IoT services or other latency-sensitive applications through a distributed architecture~\cite{fog-survey}. 

\begin{figure}[t!]
    \centering
    \includegraphics[width=\columnwidth]{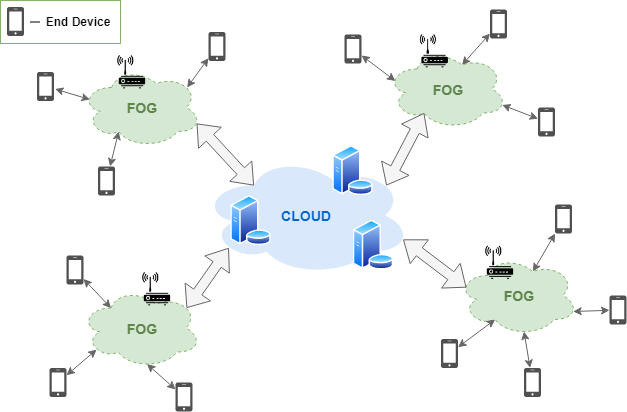}
    \caption{Cloud-Fog-Device framework.}
    \label{clofode}
\end{figure}

\subsection{Fog Computing Categories in IoT}
Fog computing consists of two categories: Cloud-Fog-Device and Fog-Device~\cite{fog-app}.

\begin{enumerate}[wide=0pt, listparindent=1.25em, parsep=0pt]
    \item \textbf{Cloud-Fog-Device.} A representation of this category is shown in Fig.~\ref{clofode}. It consists of three layers arranged in increasing order based on storage and computing capabilities.
        \begin{enumerate}
            \item \textit{Device} - An end device in an IoT network can be either mobile or fixed, depending on its application~\cite{fog-app}. Most mobile IoT devices can be either worn or carried. Some examples of heterogeneous user-oriented devices are fitness trackers, smartphones, and smartwatches. Fixed IoT devices have specific areas where they are deployed depending on their intended tasks. These types of end devices have limited energy and computing resources. They are only there to collect data~\cite{fog-energy}. Not many computations are carried out on these devices. The edges send the data to the higher layers for analysis and long-term storage.
            \item \textit{Fog} - A fog is any device that has the capability of computation, networking, and storage~\cite{fog-survey}. Some examples of these devices are switches, routers, proxy servers, bridges, and any other computing device~\cite{fog-app}. As a result, time-sensitive applications can run in fog nodes. 
            \item \textit{Cloud} - This layer is a computing and storage platform that provides various IoT applications. Cloud servers accommodate on-demand data storage and other server resources that are accessible to any device connected to the Internet~\cite{fog-app, cloud-geo}. 
        \end{enumerate}

    \item \textbf{Fog-Device} A representation of this category is shown in Fig.~\ref{fode}. It is composed of two layers with a similar arrangement in terms of storage and computing capacities. 
    
    \begin{figure}[t!]
    \centering
    \includegraphics[width=\columnwidth]{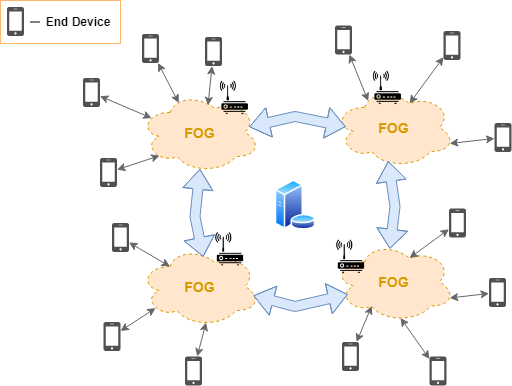}
    \caption{Fog-Device framework.}
    \label{fode}
\end{figure}

        \begin{enumerate}
            \item \textit{Fog} - In this layer, the fog nodes cooperate without the use of a cloud server to provide service to the devices~\cite{fog-app}. Each node decides on actions within a network using a distributive structure where they work as a collective unit~\cite{decentralized-fog}. The fog is the highest layer in this category. As a result, its nodes will have to handle the storing of device data without a dedicated cloud server for storage.      
            \item \textit{Device} - Similar to the Cloud-Fog-Device framework, the lowest layer is the end device. However, more of the computations are allocated by the network to the end device~\cite{decentralized-fog}. With the absence of a cloud server, more processes are required to be placed on the device to relieve the strain on the fog servers. The edges will still send the data to the fog layer for analysis and storage.
        \end{enumerate}
\end{enumerate}

Our paper focuses on the Cloud-Fog-Device as the structure of our proposed framework. This decision was attributed to the simplicity of its potential implementation compared to the Fog-Device framework. A Fog-Device network would require an algorithm that would allow better collaboration between multiple servers to achieve proper decentralization. As a result, a more complex design would be warranted. On the other hand, Cloud-Fog-Device will be easier to implement since the layers and their purposes are better defined. Also, the data flow is more straight-forward for Cloud-Fog-Device as it travels from end device to fog to cloud then back. Fog-Device is more complex because it warrants a protocol that can manage the data received from the end devices before they are presented.
 
\subsection{Open Issues for IoT- Fog Frameworks}
The addition of fog computing can improve any IoT service. However, a poorly constructed fog-based IoT network might ignore significant aspects of the network which allow it to function effectively and result in many issues.  Aspects such as power consumption, data throughput, response time, server uptime, and many more features that pertain to the network's quality of service should be taken into consideration. Ignoring these would result in a poorly constructed network. 

To determine these issues, we propose an IoT-based Urban sound sensing fog framework. Urban sound classification is a type of environmental sensing framework that focuses on categorizing the different sounds within a city area~\cite{Urbansound}. It is composed of multiple sensing nodes for recording sound and a central server for storing data and classification results. Depending on the configuration of the framework, the collected sound the classification if either executed in the end device or the server. Upon initial investigations on the feasibility of an IoT-based version of the framework, some important issues needed addressing~\cite{fog-app, hier-fog}:

\begin{itemize}
    \item \textit{High Edge Device Power Consumption.} In an urban sound sensing framework, devices transmit data to the server at intervals~\cite{ProtoLivingLab}. Each device must transmit independently and is self-sustaining. Therefore, they should be optimized before the service deploys them. However, managing each device becomes more complex as their numbers increase. One of the main issues of sensing devices is its battery life~\cite{fog-survey}. Since services deploy these devices in areas without a reliable power source, they need to account for battery capacities. Sound is being recorded continuously by this sensing framework. As a result, power becomes a challenge if the software driving the device is not optimized. Power consumption must be minimized to get the most our of each deployment. It could define the lifespan of the framework.
    \item \textit{Difficult Data Management due to Decentralization.} The end devices within a fog framework are decentralized~\cite{decentralized-fog}.  If it is improperly optimized and programmed, it could lead to inaccurate data~\cite{fog-survey}. Data precision is essential in classifying the types of sound that are being recorded by the devices. Without a proper data managing design for the end devices and the fog nodes, data traffic will be an issue for the server that will store all the results. Real-time applications such as urban sound classification will have data that will continuously be streamed to the server by multiple nodes~\cite{iot-fog-enabled, ProtoLivingLab}. If the receiving end of the server cannot handle the volume of data, then the framework will not work. 
        
\end{itemize}

High power consumption and difficult data management become more significant issues on the framework once the number of end devices and fog nodes increases. Therefore, each should be addressed to be able to achieve a scalable IoT framework.

\subsection{Proposed Solutions}
Each proposed solution is aimed to address one or both open issues and to achieve an optimized urban sound sensing IoT framework. Urban sound classification is an application that determines the type of sounds in a city using environmental sensing~\cite{Urbansound}. It is an IoT framework that monitors an area for informatic purposes~\cite{env-sound}. For urban sound classification, its focus is the different sounds of the city. However, continuous data transmissions are a requirement for an urban sound sensing framework~\cite{ProtoLivingLab}. As a result, power is a concern to end devices that have a limited supply~\cite{power-sense}. Also, cloud servers will suffer from the latency if processes are no offloaded properly to the lower layers~\cite{hier-fog}. The following are solutions proposed by this paper:
\begin{enumerate}
    \item \textit{Active Low and High Power States.} Like any sensing framework, continuous data transmissions are a concern to end devices that have a limited supply~\cite{env-sound, power-sense}. In IoT-based fog computing, power management is important to keep a standard quality of service~\cite{fog-survey}. We can optimize each device by setting each peripheral within the device to only be powered if needed. Implementing power states on the end devices could increase the efficiency of the framework. However, this could be a daunting task as the number of edges increase.   
    \item \textit{Process reallocation.}  In urban sound classification, data will be in and out of the fog and cloud servers continuously. Similar to all sensing frameworks, it will cause each server to fail once it can no longer manage the data traffic efficiently~\cite{env-sound}. However, if we can keep the incoming data at a stable and workable size, then it will be easier to manage. Transmitted data should not vary and remain small. The proposed solution aims to find the best process configuration within each layer of the framework. We can reallocate processes to achieve a balance between computation load and times~\cite{hier-fog}. With reallocation, the network can anticipate the incoming data. As a result, implementing proper server scheduling methods becomes possible. Also, with processing loads balanced, latency will be less of an issue, and we can reduce data traffic to a minimum~\cite{fog-survey}.
\end{enumerate}

\section{Proposed Framework} \label{proposal}
This section provides an overview of the design of our framework that implements our proposed solutions.

\subsection{Design Overview}
An urban sound classifier is a sensing framework that analyzes the different sounds within a city area~\cite{Urbansound}. The collected data is categorized based on a defined set of sound types~\cite{env-sound}. We chose this because it is an application that requires fog computing as an architecture. It requires multiple sensing nodes resulting in large volumes of incoming and outgoing data. To implement this framework, we need to consider real-time data processing and big data management. Ideally, fog computing should be capable of these due to its load balancing and node management features ~\cite{fog-priv, fog-med}. However, as mentioned in the previous section, fog computing was running into issues in high power consumption in end devices and data management in servers. These could prevent fog computing from being a reliable option for this IoT-based urban sound sensing framework. This paper proposes active low and high power states and process reallocation to address these concerns.  

\begin{figure*}[t!]
    \centering
    \includegraphics[width=0.9\textwidth]{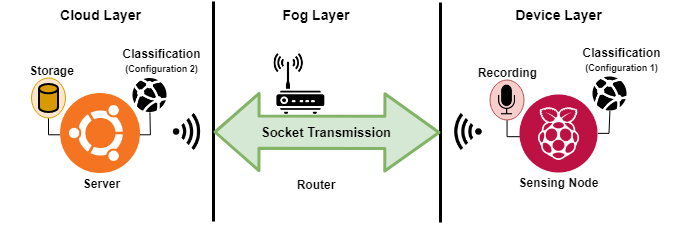}
    \caption{Framework setup and layer hierarchy.}
    \label{fms}
\end{figure*}

We created an implementation of this urban sound classifying framework. It uses a star topology with a central server and end devices. Also, the design makes use of a remote desktop as the server. This server runs on an Ubuntu 16.04 operating system. Each end device is a Raspberry Pi 3 Model B loaded with a Raspbian-Jesse operating system (OS). Then, we installed the OS through the NOOBS software provided by the Pi homepage. Connecting the devices to the server is a wireless client-server socket setup. We used an STM32 NUCLEO-64 board with an attached X-NUCLEO-CCA02M1 expansion board. This hardware choice made it easier to record sound and convert it into formatted data. Python 3.6 was chosen to be the programming language for the scripts to run the required libraries for the classifier and the digital microphone manager. The server will be using Ubunutu 16.04 as its operating system with a wireless network router as its fog. This hierarchy and the setup of our framework is shown in Fig.~\ref{fms}. 

Raspberry Pi 3s devices were selected due to their modularity and affiliation towards rapid prototyping. Our framework requires multiple nodes that are identical in functionality. Pis can also be easily reprogrammed and adapted in situations when immediate changes to any of our scripts are needed. It is a strong development tool that allows us to incorporate other technologies such as the STM32 NUCLEO-64 board without any difficulties due to its flexibility. Also, Pis are very low powered which is an advantage in our testing when it comes to us requiring multiple nodes that are running at the same time.

\subsubsection{Configuration setup}
\begin{figure}[t!]
\centering
\subfloat[Section setup.]
{   \includegraphics[width=\columnwidth]{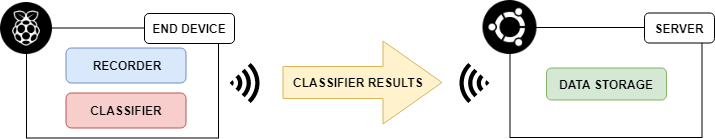}\label{conf1setup}}\\
\subfloat[Process allocation.]
{    \includegraphics[width=\columnwidth]{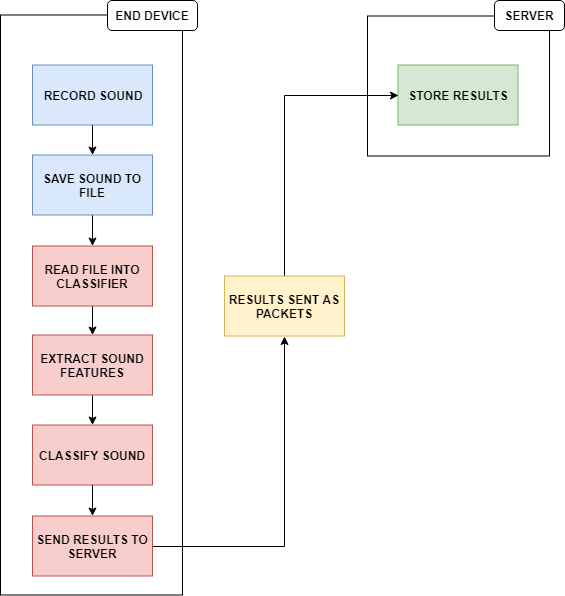}\label{conf1alloc}}
\caption{Configuration 1 - Design and logic details.}
\label{conf1}
\end{figure}

\begin{figure}[t!]
\centering
\subfloat[Section setup.]
{   \includegraphics[width=\columnwidth]{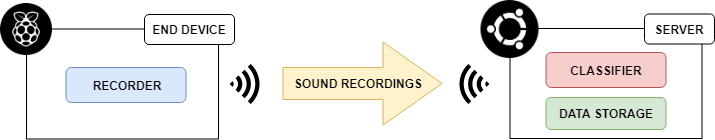}\label{conf2setup}}\\
\subfloat[Process allocation.]
{    \includegraphics[width=.8\columnwidth]{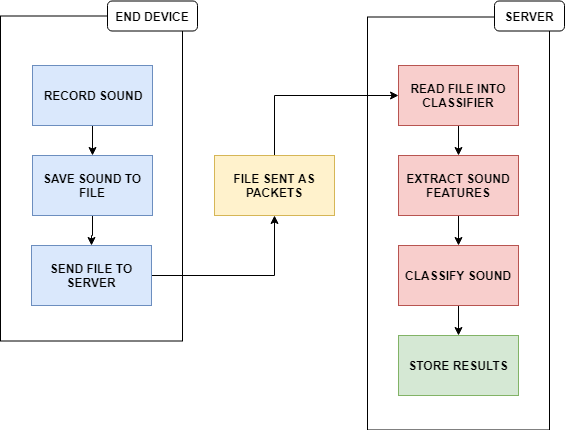}\label{conf2alloc}}
\caption{Configuration 2 - Design and logic details.}
\label{conf2}
\end{figure}

We created two initial configurations for the initial testings. These derived frameworks will be tested based on end-device power consumption and server latency to check their performances. Then, we will compare their results to a third, proposed configuration. It will be formulated based on our proposed solutions to determine their feasibility in creating a more scalable IoT-framework. Our program needs to be flexible with sections and processes that can be reallocated along with the network layers to support these configurations. In total, there are three configurations in this paper:

\begin{enumerate}
    \item \textbf{Configuration 1} -  The first configuration will represent a framework that relies heavily on the edge devices and the fog. The division of the sections within it is shown in Fig.~\ref{conf1setup}. In this configuration, we allocated most of the processes at the edge of the network in the end devices. Its allocation within the program is shown in Fig.~\ref{conf1alloc}.
    \item \textbf{Configuration 2} - The second configuration will represent a framework that relies heavily on the cloud server. The division of the sections within it is shown in Fig.~\ref{conf2setup}. In this configuration, we will reallocate the movable processes to the cloud server. Its reallocation within the program is shown in Fig.~\ref{conf2alloc}.
    \item \textbf{Proposed Configuration} - The third or our proposed configuration will be an implementation of the solutions that we presented. We will use the active low and high power states as well as resource reallocation to create a better configuration. Within each is an arrangement of the program that will carry out the different processes in the framework. 
\end{enumerate}

For the first two configurations, the program is initially broken down into two major sections: the sound recorder and the sound classifier. 

\subsubsection{Sound Recorder}
The sound recording section is composed of the digital microphone and a Python script to drive it. We used a Python library called PySound to calibrate and read from the mic. A commercial digital Micro-Electro-Mechanical Systems (MEMS) microphone was used to make the interfacing and prototyping easier. The microphone setup was set to be their default, out-of-the-box configurations (i.e.16 kHz sampling rate,  single-channel, 16-bit resolution, 4096-byte frame size) to keep the design simple. 

First, we initialize the PyAudio library within the script used to record the sound. Then, PyAudio will start the recording by creating an audio stream. During the recording process, the data read in by the Pi from the microphone is broken down into size-defined data frames. The selected frame size was 4096 bytes per frame. Upon reaching the recording length, the library closes the stream. Lastly, the resulting data is stored by the script into a Waveform Audio File Format (WAV) file.

\subsubsection{Sound Classifier}
The sound classifying section is composed of a dataset, a classify, and a script to train and run the classifier. We chose urban sound classification due to the insurgence of interest in smart cities. The dataset used is the UrbanSound8K dataset~\cite{Urbansound}. The UrbanSound dataset is a collection of 1302 full-length recordings with each urban sound labeled according to the sound occurrence and salience annotations. Each sound clip in the dataset was obtained from an online free sound file provider. The 8K iteration of the dataset cuts each sound clip into 4-second segments. They cut the sound clips because of a listening test conducted. This listening test showed that 4 seconds was the best clip duration which yielded an accuracy of 82\%. Within the UrbanSound8K dataset is a collection of 10 low-level classes: air conditioner, car horn, children playing, dog barking, drilling, engine idling, gunshot, jackhammer, siren and street music. We based the classifier on~\cite{Urbansound}. The script makes use of a combination of Librosa and Tensorflow to carry out the extraction and classification.

Feature extraction is a part of pre-processing data before it gets used to train and test the classifier model. This process obtains these features as vectors that represent specific aspects of the sound clip~\cite{feature-ext}. The script makes use of the sound processing library called Librosa to extract these features. This library loads in each file from the dataset and converts them into feature vectors. Then, the conversion is carried out using feature extraction methods provided by Librosa. The script that we used then saves each feature vector into text files that are parsed back into the program during training and tests. We used Tensorflow to build and train the model. It is a neural network composed of 2 hidden layers, each having 280 and 300 nodes respectively. After processing the dataset, the script trains the neural network under 3000 epochs. The training process makes use of the dataset, which we split 70-30; 70\% of the sound files to train the classifier while the remaining 30\% for testing and verification. After multiple iterations, the best accuracy was 85\% using a training epoch of 5000 and a learning rate of 0.1.

\section{Testing and Evaluation} \label{exp}
This section consists of an overview of the testing setups, followed by a discussion on the end device power consumption and server latency tests.

\subsection{Testing Overview}
As mentioned in the overview, we have three configurations that will be used by our tests. The first two configurations will model the default setups for fog and cloud architectures, respectively. The third configuration will implement our proposed active low and high power states and process reallocation technique. We will focus on metrics that determine the end device power consumption and server data management capabilities of each configuration.

We determined high power consumption among end devices and data management as issues in an IoT-based fog framework. As a result, we introduced active low and high power states and process reallocation. We can make the network run only parts of the architecture by defining low and high power states. We programmed each Pi to use only the necessary peripherals for each process. We toggled its wireless component so that it would only run when it transmits data between the device and the server. As a result, we can isolate the main sections of the design as the only significant power consumers.  

There are two main sections in the design; sound recorder and sound classifier. We determined which of these sections can be moved between the end device and server to know which parts can conserve energy. The recording process needs to stay within the Pi because it is what drives the digital microphone. However, the sound classifier can either be placed on the server or remain within the Pi. We created a program that can be divided into these sections so we can effectively reallocate the processes. 

\begin{figure}[t!]
    \centering
    \includegraphics[width=\columnwidth]{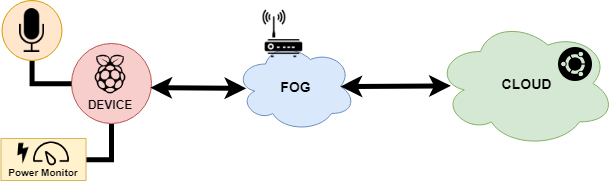}
    \caption{Power consumption test setup.}
    \label{powconsetup}
\end{figure}

\subsection{Power Consumption Tests}
\begin{figure}[t!]
    \centering
    \includegraphics[width=\columnwidth]{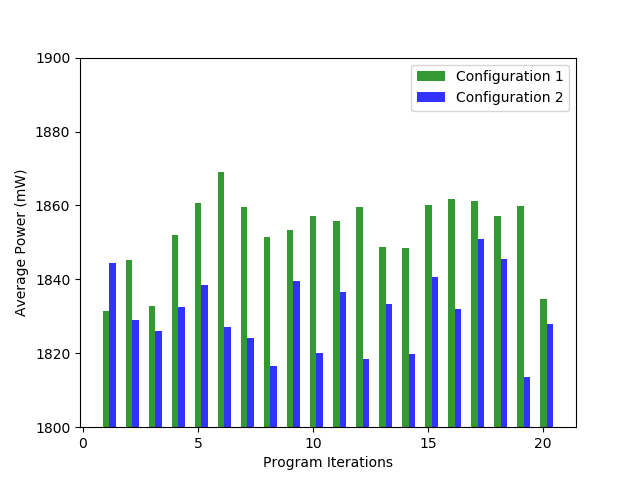}
    \caption{Power consumption of each configuration in 20 iterations.}
    \label{locserpow}
\end{figure}

We conducted our first test to check which of the two configurations between the classifier placement requires less power. A sample setup of the test is in Fig.~\ref{powconsetup}. The first configuration executes both the sound recorder and classifier within the device. The second configuration moves the classifier from the device to the server. The testing setup has a node record the sound for 10 seconds and has it classified either within the same Pi or the server. Then, the power consumed is measured and compared. We executed this setup 20 times, and the results are shown in Fig.~\ref{locserpow}. According to the results, the first configuration had an average power consumption of 1852.00~mW while the second configuration had 1830.54~mW. These values show no significant difference in the two configurations. 

In addition, the experiment was conducted overnight for 10~hours to obtain a reasonable dataset. During this time, the power readings were consistent without showing any signs of overheating such as significant spikes in power or long periods of high power usage on the side of the Raspberry Pis or the server. As a result, the conducted power consumption tests also further support the feasibility of the framework of being deployed in a real environment.

\begin{figure}[t!]
    \centering
    \includegraphics[width=\columnwidth]{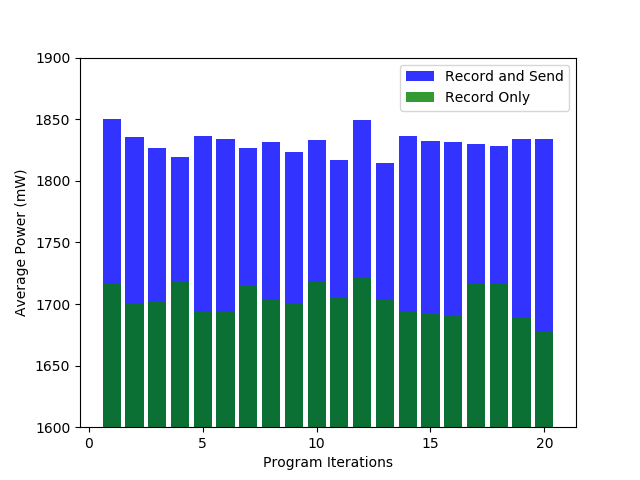}
    \caption{Difference in power consumption with and without sending the data to the server after recording.}
    \label{recpow}
\end{figure}

We chose to add another test to check the distribution of the measured power consumption within each configuration. In this test, we measured the effects of wireless transmission to the power consumption of the framework by testing the sound recording section. The outcome of the test within 20 iterations is shown in Fig.~\ref{recpow}. Averaging the results showed a power difference of 127.54~mW. For every iteration, the first configuration transmits around 130~kilobytes (KB) of data while the other configuration only transmits 4~bytes (B) being the size of a single integer. As a result, data size becomes a significant variable for wireless transmission. The first configuration sends a whole sound file to the server for classification. The second configuration only sends the results of the classifier. 

These two configurations may have yielded the same power consumption, but it was because of different reasons; processing power and data size. The first configuration takes all the processing load by executing both recording and classifying portions of the framework. However, the second configuration demands more power within the wireless transmission to move the sound data from the end device to the server. Having smaller transmitted data sizes saves energy, but it is canceled out by processing demands of classifying at the edge of the network. 

We measured the runtime of each configuration in 20 iterations. The results were an average runtime of 57.77 and 16.42 seconds for the first and second configuration, respectively. Therefore, this time difference points out that the measured power consumption is due to the placement of the classifying process. Also, the runtime results point out the disparity in the execution times of the first two configurations. There is no significant difference in power consumption between the configurations. The power drawn by running the classification in the end device is offset by the power required to transmit the sound files to the cloud for classification. However, the time differences points out at advantage of having the classifier in the server. It improves the overall time it takes to carry out the whole sound classification process.

Finally, we tried to create a configuration that balances the processing load and the size of the data transmitted. The classifier is composed of two subsections; the feature extraction and the actual classification. Before a sound file is classified, its feature vectors are first extracted. We divided this section into two and having the feature extraction process moved to the end device. As a result, the data size changed from around 130~KB to a constant 1.6~KB. This consistency is due to the feature vectors predefined to have a dimension for every sound clip. 

\begin{figure}[t!]
\centering
\subfloat[Section setup.]
{   \includegraphics[width=\columnwidth]{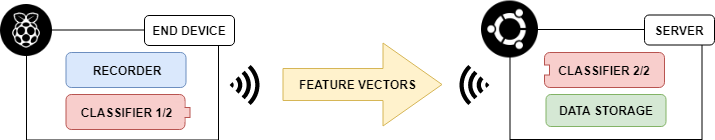}\label{conf3setup}}\\
\subfloat[Process allocation.]
{    \includegraphics[width=\columnwidth]{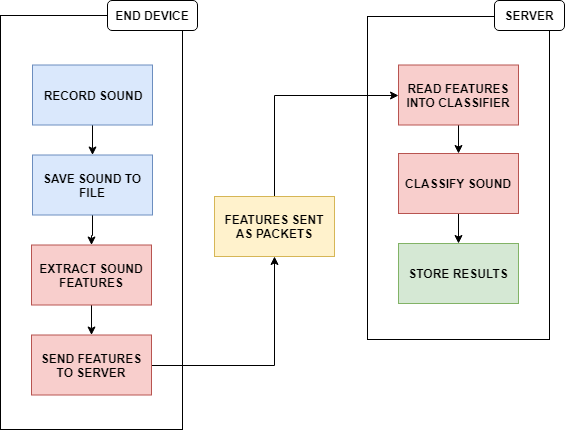}\label{conf3alloc}}
\caption{Proposed configuration - Design and logic details.}
\label{conf3}
\end{figure}

Though we added more processes to the end device, it was not as much as the first configuration. Also, at the cost of the addition, the data transmission has become more constant and stable in case the design required longer recording times. The resulting setup division for the proposed configuration is shown in Fig.~\ref{conf3setup} and the process reallocation is shown in Fig.~\ref{conf3alloc}. We tested this configuration under the same setup resulting in an average power consumption of 1786.86~mW. Comparing these results to the previous ones, it shows that this is an improved configuration.

\subsection{Latency Tests}
The next concern is the server data management. Once the number of nodes within the framework increase, we need to examine if it will be able to handle the incoming data. This test took the three configurations used in the power consumption test and verifies their viability in data management. We used latency to measure each of their effectiveness in managing data. Latency is the time it takes a data packet to finish transmitting from the device to the server. The experimental setup implements a star topology with multiple nodes loaded with the same configuration. Each node sends data to the server. The server then takes each packet and measures the time between the beginning and the end of the transmission. 

\begin{figure}[t!]
    \centering
    \includegraphics[width=.7\columnwidth]{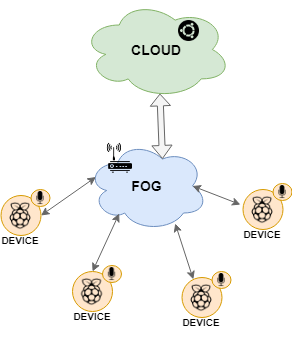}
    \caption{Latency test setup.}
    \label{latsetup}
\end{figure}

\begin{figure}[t!]
    \centering
    \includegraphics[width=\columnwidth]{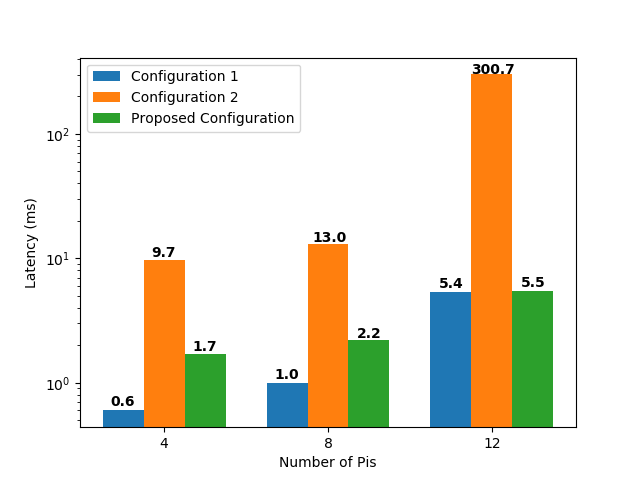}
    \caption{Latencies of configurations in a star topology setup with 4, 8, and 12 Pis.}
    \label{latlocserhyb}
\end{figure}

A sample setup of the latency test is shown by Fig.~\ref{latsetup}. The test setup uses a round-robin type of schedule. All nodes are listed based on their given IP address. The implemented scheduler goes through this list and allows the selected device connection one at a time. The latency of each configuration was tested by creating networks that had a varying number of nodes. We programmed each device to transmit data to the server for the same duration as the power consumption tests. The data was sampled for 10 iterations. The results of the tests where each configuration consists of a network with 4, 8 and 12 nodes are shown in Fig.~\ref{latlocserhyb}. The results show that as the number of nodes increased, both the first and our proposed configuration were able to handle the increase. However, as the number of nodes reached 12, the latency of the second configuration jumps to 300 milliseconds. This jump in value depicts significant packet loss that shows how this configuration cannot handle the current setup, which had 12 nodes. 

Wireless interference is a common reason for such a significant difference in latency. However, the tests for each configuration were conducted with the same components and environment. If wireless interference was present, it would have affected all the tests and not just the second configuration. Thus, latency is more attributed to the nature of the data being transmitted and the time spent to compute the results in the server. Overall, the results of our tests show that our proposed configuration was the best out of the three.

\section{Conclusion} \label{conc}
Fog computing with IoT networks runs into power consumption and data management issues. Due to these, scalability becomes harder to implement when using fog computing. Urban sound classification is a sensing framework that benefits from fog computing. This paper proposes process reallocation and active-low and high states to create a more scalable framework. Our tests resulted in a configuration that addresses each issue effectively. 

In terms of power consumption, data size and processor load is a significant variable. The proposed configuration shows a balance between a constant data packet size and a logically allocated processor load. For data management, latency is crucial for a growing network while data speed is also important. It dictates how a server is capable of managing its incoming data. Both the first and our proposed configuration proved to be scalable based on their latency test results. However, by taking into consideration all of the test results from both issues, the proposed configuration that implemented optimization techniques proved to be better overall.

The first configuration showed to be more device heavy while the second was more server heavy. Each is a representation of the default fog and cloud computing architecture, respectively. As a result, we created a third configuration which is a hybrid of both architectures. It shows the importance of balance between the two extremes focusing on more adaptive design.  With this configuration, proper processor load balancing can improve an end device's power consumption. Also, transmitted data size management helps in a server's data management. Therefore, process allocation and power management through active-low and high states can improve the scalability of an IoT-based urban sound sensing fog framework. 

\bibliography{fogbib}
\end{document}